# Novel Attributes of a
# Dual Material Gate Nanoscale Tunnel Field Effect Transistor

Sneh Saurabh and M.Jagadesh Kumar, *Senior Member, IEEE,*

Department of Electrical Engineering, Indian Institute of Technology, Hauz Khas, New Delhi 110016

email: s.saurabh@gmail.com and mamidala@ieee.org

*Abstract*— In this paper, we propose the application of a Dual Material Gate (DMG) in a Tunnel Field Effect Transistor (TFET) to simultaneously optimize the on-current, the off-current and the threshold voltage, and also improve the average subthreshold slope, the nature of the output characteristics and the immunity against the DIBL effects. We demonstrate that if appropriate work-functions are chosen for the gate materials on the source side and the drain side, the tunnel field effect transistor shows a significantly improved performance. We apply the technique of DMG in a Strained Double Gate Tunnel Field Effect Transistor with a high-k gate dielectric to show an overall improvement in the characteristics of the device along with achieving a good on-current and an excellent average subthreshold slope. The results show that the DMG technique can be applied to TFETs with different channel materials, channel lengths, gate-oxide materials, gate-oxide thicknesses and power supply levels to achieve significant gains in the overall device characteristics.

*Index Terms*— Dual Material Gate, Tunnel Field Effect Transistor, on-current, off-current, threshold voltage, strain.

# I. INTRODUCTION

Tunnel Field Effect Transistors (TFETs) exhibit an excellent subthreshold swing and a very low leakage current and hence are being actively investigated for future low power CMOS applications [1-20]. However, TFETs suffer from a low on-current ($I_{ON}$) and therefore, various techniques to improve the $I_{ON}$ in a TFET have been suggested [4-14]. The mechanism of current transfer and the saturation in the output characteristics in a TFET is quite different from a conventional MOSFET [16]. As a result, TFETs often exhibit delayed saturation in the output characteristics. This can potentially be detrimental for CMOS applications, and therefore the nature of the output characteristics must be carefully considered while designing TFETs. Also, the dependence of the drain current on the drain voltage in a TFET is different from a conventional MOSFET [16, 18]. Strong DIBL effects is sometimes manifested in a TFET and this can severely limit the utility of the device [18]. Therefore, to employ TFET for low power CMOS applications, it is desirable that in addition to improving the on-current and the subthreshold swing, the overall device characteristics of the TFET be improved. In this paper, we propose the application of a Dual Material Gate (DMG) [21-23] in a Double Gate Tunnel Field Effect Transistor (DGTFET) and demonstrate using 2D device simulations that by engineering the work functions of the dual gates, it is possible to simultaneously optimize the on-current ($I_{ON}$), the off-current ($I_{OFF}$) and the threshold voltage ($V_T$), and also improve the average subthreshold slope ($SS_{AVG}$), the nature of the output characteristics and the immunity against DIBL effects.

It has already been demonstrated in the previous works that a conventional DGTFET suffers from an unacceptably low $I_{ON}$ [4-10, 20]. The DMG technique alone cannot bring the $I_{ON}$ of a DGTFET to that of the MOSFET levels. It has also been shown that when the channel is composed of a smaller band-gap material, the $I_{ON}$ in a TFET improves dramatically [10-14]. Therefore, we demonstrate the application of the DMG technique in a TFET composed of a small band-gap material so that the $I_{ON}$ is good enough for practical purposes. In ref. [10], the Strained DGTFET (SDGTFET) has been demonstrated to improve the $I_{ON}$ and the subthreshold swing. SDGTFET is structurally same as a DGTFET, except that the silicon body is composed of SiGe free sSOI. The strain in an SDGTFET is controlled by the Ge mole-fraction (x) that has been used to fabricate the sSOI [25-27]. Since the introduction of strain decreases the band-gap of the silicon, an SDGTFET shows a remarkably improved device performance [10]. In ref. [5], it has been shown that a high-k gate dielectric improves the coupling between the gates and the tunnel junction resulting in an improved performance of a DGTFET. Therefore, we demonstrate the application of the DMG technique in an SDGTFET with a high-k gate dielectric to not only achieve an improved $I_{ON}$ and $SS_{AVG}$ but also improve the overall performance of the device. It is also demonstrated that, using the additional handle of the DMG, the scalability of the TFET can be extended below 20 nm channel lengths. Finally, considering the potential application of a TFET as a low power device, the suitability of DMG-SDGTFET is studied at $V_{DD} = 0.5$ V [20].

The rest of this paper is organized as follows: Section II describes the structure of a DMG-DGTFET and the simulation model used in this study. Section III presents simulation results for a DMG-DGTFET and demonstrates the advantages of using a DMG. In this section the channel length of the device is taken as 50 nm since the physical phenomenon are easier to appreciate at comparatively larger device dimensions. Section IV demonstrates the improvement in the characteristics of the device due to DMG in an SDGTFET. In this section the channel length is taken to be 25 nm. The analysis of the DMG-SDGTFET has been done for $V_{DD}$ = 1.0 V followed by $V_{DD}$ = 0.5 V. Finally, Section V draws important conclusions out of this study.

.

## II. DEVICE STRUCTURE AND SIMULATION MODEL

Fig. 1 shows the cross-sectional view of the proposed DMG-DGTFET in which both the top and the bottom gates are composed of materials with two different work functions. We refer to the gate closer to the source as the tunnel-gate and the one closer to the drain as the auxiliary-gate. The fabrication of a DMG-DGTFET could be done using the techniques reported for fabricating a DMG-FET [21,22]. All the simulations have been carried out using ATLAS version 1.12.1.R [24]. Since the tunneling process is non-local, spatial profile of the energy bands and the band gap narrowing effects are included [24]. We have used non-local tunneling model in this study and validated our simulation model using ref. [5]. We have taken the source and drain doping profiles as abrupt throughout our simulations as in earlier works [4, 5, 10, 12, 15]. The simulation model used in this paper, including the tunneling model and the model for strained silicon, has also been used in ref. [10, 15].

The device parameters used are: source doping = $10^{20}/cm^3$, drain doping = $5x10^{18}/cm^3$, body doping = $10^{17}/cm^3$, silicon body thickness ($t_{si}$) = 10 nm and gate oxide thickness ($t_{ox}$) = 3 nm, channel length (L) = 50 nm, tunnel gate length ($L_{tunn}$) = 20 nm and auxiliary gate length ($L_{aux}$) = 30 nm.

## III. DUAL MATERIAL GATE DGTFET

First, we analyze the impact of the work-function of the auxiliary-gate ($\Phi_{aux}$) on the $I_{OFF}$ in a DMG-DGTFET. Fig. 2(a) and 2(b) show the change in the band diagram of the device as $\Phi_{aux}$ is increased, keeping $\Phi_{tunnel}$ fixed at 4.0 eV. In the off-state (Fig. 2(a)), as $\Phi_{aux}$ is increased, the tunneling width increases and the band-overlap decreases on the source side leading to a considerable reduction in the off-state tunneling probability. However, at higher $\Phi_{aux}$ (> 4.4 eV), band overlap begins to appear on the drain side and if the tunneling junction width on the drain side becomes small (as in the case of $\Phi_{aux}$ = 4.8 eV), tunneling can occur in the off-state. In the on-state (Fig. 2(b)), the increase in $\Phi_{aux}$ does not change the band diagram significantly. The band overlap begins to decrease when $\Phi_{aux}$ > 4.4 eV. Since the reduction in band overlap is over that section of the band-diagram where tunnel-width is already large, the tunneling probability is not affected significantly. Fig. 2(c) shows the change in the transfer characteristics as $\Phi_{aux}$ is increased from 4.0 - 4.8 eV. As expected, the $I_{OFF}$ (calculated at $V_{GS}$ = 0.0 V and $V_{DS}$ = 1.0 V) reduces by more than 3 orders of magnitude when $\Phi_{aux} \geq$ 4.4 eV. However, when $\Phi_{aux} \geq$ 4.8 eV, greater tunneling occurs on the drain side increasing $I_{OFF}$ again. The $I_{ON}$ (calculated at $V_{GS}$ = $V_{DS}$ = 1.0 V) reduces by not more than 15% when $\Phi_{aux}$ is increased from 4.0 - 4.4 eV.

Next, we analyze the impact of the tunnel-gate work-function ($\Phi_{tunnel}$) on the characteristics of a DMG-DGTFET. Fig. 3(a) and 3(b) show the change in the band diagram of the device as $\Phi_{tunnel}$ is decreased, keeping $\Phi_{aux}$ fixed at 4.4 eV. In the off-state (Fig. 3(a)), there is no band-overlap on the source side, when $\Phi_{tunnel}$ is reduced to 4.0 eV, and hence the $I_{OFF}$ is expected to be quite low. In the on-state (Fig. 3(b)), with the reduction in $\Phi_{tunnel}$, the band-overlap increases and the tunneling width decreases, leading to a significant increase in the tunneling probability on the source side. Fig. 3(c) shows the change in the transfer characteristics of the DMG-DGTFET as $\Phi_{tunnel}$ is decreased from 4.8 - 4.0 eV. As expected, the $I_{ON}$ increases by around 4 orders of magnitude and the $V_T$ (computed as $V_{GS}$ when the drain current reaches $1x10^{-7}$ A/μm) reduces to 0.5 V at $\Phi_{tunnel}$= 4.0 eV. The $I_{OFF}$ is below 1 fA/μm for all values of $\Phi_{tunnel}$.

The above analysis shows that $\Phi_{aux}$ and $\Phi_{tunnel}$ can be adjusted for a better $I_{OFF}$, $I_{ON}$ and $V_T$ trade-offs. Since there is a large off-state tunneling when $\Phi_{tunnel} < 4.0$ eV (on the source side) or $\Phi_{aux} > 4.8$ eV (on the drain side), we have optimized $\Phi_{aux}$ and $\Phi_{tunnel}$ within 4.0 - 4.8 eV. Fig. 3(c) shows that when we decrease $\Phi_{tunnel}$, the $V_T$ reduces and the $I_{ON}$ increases. Since in general, DGTFET suffers from a low $I_{ON}$ and high $V_T$, it is desirable to adjust the device parameters to obtain a maximum $I_{ON}$ and a minimum $V_T$. Therefore, we choose $\Phi_{tunnel}$ to be the lowest possible value i.e. 4.0 eV. Fig. 2(c) shows that when we increase $\Phi_{aux}$, the $I_{OFF}$ decreases. However, making $\Phi_{aux} > 4.4$ eV reduces the $I_{ON}$ without significantly improving the $I_{OFF}$. Therefore, 4.4 eV is chosen as the optimal value for $\Phi_{aux}$. There are several metal candidates that can be engineered to obtain the desired $\Phi_{aux}$ (4.4 eV) (e.g.: W, Ta, and Mo) and $\Phi_{tunnel}$ (4.0 eV) (e.g.: Mo, Ni-Ti and Sc) [28-30]. Next, we find an optimum value of the tunnel gate length ($L_{tunn}$), for the given $\Phi_{aux}$, $\Phi_{tunnel}$ and L (50 nm). Fig. 4 shows that the $I_{OFF}$ decreases and the $I_{ON}$ remains unchanged as $L_{tunn}$ is decreased up to 20 nm. However, there is no significant impact on the $I_{OFF}$ below 20 nm. Therefore, we have chosen $L_{tunn}$= 20 nm.

The transfer characteristics in Fig. 5 show that the DMG-DGTFET has simultaneously a high $I_{ON}$ and a low $I_{OFF}$ which cannot be achieved using an SMG-DGTFET. It may be noted that the ratio $I_{ON}/I_{OFF}$ in an SMG-DGTFET can be improved by adjusting $\Phi_m$ as demonstrated in ref. [10]. It is found that, at approximately $\Phi_m = 4.2$ eV, the best possible value of $I_{ON}/I_{OFF}$ (around $3x10^9$) is achieved in the given SMG-DGTFET. The ratio of $I_{ON}/I_{OFF}$ is in the range of $1x10^{10}$ for a DMG-DGTFET. The improvement in the ratio of $I_{ON}/I_{OFF}$ in a DMG-DGTFET is, to some extent, a consequence of the improvement in the average subthreshold slope ($SS_{AVG}$). The $SS_{AVG}$ is an important parameter for a TFET and is computed as:

$$SS_{AVG} = \frac{(V_T - V_{OFF})}{\log(I_{VT}) - \log(I_{VOFF})}$$

where $V_T$ is the threshold voltage, $V_{OFF}$ is the gate voltage from which the drain current starts to take off (as shown in Fig. 5), $I_{VT}$ is the drain current at $V_{GS} = V_T$ and $I_{VOFF}$ is the drain current of the device at $V_{GS} = V_{OFF}$ (as shown in Fig. 5) [5,31]. The $SS_{AVG}$ improves from 73 mV/decade in an SMG-DGTFET to 58 mV/decade in the DMG-DGTFET. It is worth mentioning that manipulation of the $\Phi_m$ in the SMG-DGTFET just shifts the transfer characteristics parallel to the X-axis and cannot improve the $SS_{AVG}$. However, in a DMG-DGTFET $\Phi_{aux}$ and $\Phi_{tunnel}$ provides an additional handle to manipulate different sections of the transfer characteristics, as shown using energy band-diagrams in Fig. 2 and Fig. 3, and thereby improve the $SS_{AVG}$. Fig. 6 shows the normalized output characteristics with respect to the drain current at $V_{DS} = 1$ V. The saturation voltage ($V_{DSAT}$) is computed as the $V_{DS}$ required to make $I_D$ reach 95 % of the $I_D$ at $V_{DS} = 1$ V. For $V_{GS} = 0.7$ V, $V_{DSAT}$ reduces from 0.94 V for an SMG-DGTFET to 0.74 V for a DMG-DGTFET. The saturation voltage in a TFET corresponds to the disappearance of the inversion charge in the channel [16]. The electron concentration along a horizontal cutline close to the gate at different $V_{DS}$ is shown in Fig. 7. Since $\Phi_{aux}$ > $\Phi_m$, the inversion charge density in the channel on the drain side is lower in the DMG-DGTFET compared to the SMG-DGTFET. Therefore, disappearance of the channel charge is facilitated in the DMG-DGTFET and a quicker onset of saturation is observed. It may be noted that a low $\Phi_m$ in a DGTFET results in a delayed saturation in the output characteristics. However, increasing $\Phi_m$ would result in an undesirable decrease in $I_{ON}$ and an increase in $V_T$. In a DMG-DGTFET, a low $V_{DSAT}$ and a high $I_{ON}$ is attained simultaneously by using a high $\Phi_{aux}$ and a low $\Phi_{tunnel}$. Hence, the DMG-DGTFET exploits the advantages of both the low and the high work function gate to attain an overall improvement in the device characteristics. Fig. 8(a) and Fig. 8(b) shows the transfer characteristics of the SMG-DGTFET ($\Phi_m = 4.2$ eV) and the DMG-DGTFET respectively. The difference in $V_{GS}$ at a constant drain current for the transfer characteristics of $V_{DS} = 1.0$ V and $V_{DS} = 0.1$ V (as shown in Fig. 8) is a manifestation of DIBL in a DGTFET [18]. The difference in $V_{GS}$ that is marked in Fig. 8 is found to be 0.215 V for the SMG-DGTFET and 0.131 V for the DMG-DGTFET. Therefore, the DMG-DGTFET is more immune against DIBL effects.

## IV. DUAL MATERIAL GATE SDGTFET

Though the results presented in the previous section show a significant improvement in the characteristics of a DGTFET due to the application of the DMG, the $I_{ON}$ of the DMG-DGTFET is around 10 µA/µm which is unacceptably low for the CMOS applications [20]. One of the most effective techniques to improve the $I_{ON}$ is to choose a lower band-gap material for the channel [10-14,20].Therefore, we choose a s-SOI based SDGTFET (Ge mole fraction x = 0.2) in this section and also make the following modifications to the device that was studied so far: a) $HfO_2$ (k = 21) with $t_{ox} = 2$ nm is used as the gate oxide to further boost the $I_{ON}$ [5] b) channel length L is taken as 25 nm so that it is more realistic with the future applications. The detailed description of the impact of strain on the material properties and the simulation parameters of the silicon can be found in ref. [10,15]. It should also be noted that, as in the previous works [5,12,17,18,31], we have not considered the gate leakage through the thin $HfO_2$ gate oxide. Using the methodology outlined above for the DGTFET, the optimum device parameters for the $I_{ON}$, $I_{OFF}$, and $V_T$ trade-offs in a DMG-SDGTFET are found to be:

$\Phi_{tunnel} = 4.3$ eV, $\Phi_{aux} = 4.6$ eV and $L_{tunn} = 12$ nm. The transfer characteristics obtained for SMG-SDGTFET and the DMG-DGTFET are shown in Fig. 9. The DMG-SDGTFET is able to deliver a high $I_{ON} = 351$ µA/µm, $I_{OFF} < 1$ fA/µm, the ratio of $I_{ON}$ / $I_{OFF}$ is around $3 \times 10^{12}$ and $V_T = 0.20$ V. The $SS_{AVG}$ improves from 34 mV/decade in an SMG-SDGTFET to 21 mV/decade in a DMG-SDGTFET. The $V_{DSAT}$ at $V_{GS} = 0.7$ V reduces from 0.75 V in the SMG-DGTFET ($\Phi_m = 4.3$ eV) to 0.65 V in the DMG-DGTFET. The DIBL (computed as the difference in $V_{GS}$ at $I_D = 1$ nA/µm for the transfer characteristics of $V_{DS} = 0.1$ V and $V_{DS} = 1.0$ V) also improves from 86 mV in an SMG-SDGTFET ($\Phi_m = 4.6$ eV) to 43 mV in a DMG-SDGTFET. The above results show that an excellent device characteristic is achieved in a DMG-SDGTFET.

Fig. 10 shows the change in $V_T$ with L for an SMG-SDGTFET and a DMG-SDGTFET. For an SMG-SDGTFET there is no change in $V_T$ up to around 20 nm. However, when L < 20 nm, the SMG-SDGTEFT does show significant $V_T$ roll-off [10,17]. A DMG-SDGTFET also shows no change in $V_T$ up to 20 nm when the $L_{tunn}$ is fixed to 12 nm. However, for L ≤ 20 nm, $L_{tunn}$ can be adjusted ($L_{tunn} = 12$ nm for L = 20 nm, $L_{tunn} = 10$ nm for L = 15 nm and $L_{tunn} = 7$ nm for L = 10 nm) to alleviate the problem of $V_T$ roll-off as shown in Fig. 10. Therefore, DMG provides an additional handle to extend the scalability of the TFET. The $I_{ON}$ of both an SMG-SDGTFET and a DMG-SDGTFET is not significantly affected by scaling.

Since Tunnel FETs are being actively investigated as enablers of future logic circuits operating with a $V_{DD} < 0.5$ V, we analyze the suitability of the DMG-SDGTFET at $V_{DD} = 0.5$ V [20]. Since low power applications impose a stricter requirement on the subthreshold swing, we choose an SDGTFET with higher Ge mole fraction (x = 0.5) [10]. The optimal device parameters for DMG-SDGTFET with x=0.5, $V_{DD} = 0.5$ V, L =

25 nm and HfO$_2$ gate oxide with t$_{ox}$ = 2 nm are: $\Phi_{tunnel}$ = 4.5 eV, $\Phi_{aux}$ = 4.7 eV and L$_{tunn}$ = 10 nm. Fig. 11 shows the comparison of the transfer characteristics of the DMG-SDGTFET with SMG-SDGTFETs at V$_{DS}$ = 0.5 V. The important electrical parameters for the DMG-SDGTFET are: I$_{ON}$ = 35 μA/μm, I$_{OFF}$ < 1 fA/μm, ratio of I$_{ON}$ / I$_{OFF}$ is around 3x10$^{11}$ and V$_T$ = 0.16 V. The SS$_{AVG}$ improves from 28 mV/decade in the SMG-SDGTFET to 16 mV/decade in the DMG-SDGTFET. The V$_{DSAT}$ at V$_{GS}$ = 0.35 V decreases from 0.44 V in the SMG-SDGTFET ($\Phi_m$ = 4.5 eV) to 0.35 V in the DMG-SDGTFET. The DIBL (computed as the difference in V$_{GS}$ at I$_D$ = 1 nA/μm for the transfer characteristics of V$_{DS}$ = 0.05 V and V$_{DS}$ = 0.5 V) also improves from 100 mV in the SMG-SDGTFET ($\Phi_m$ = 4.6 eV) to 62 mV in the DMG-SDGTFET. The excellent SS$_{AVG}$ of the DMG-SDGTFET shows that it can be one of the suitable candidates for future low power CMOS applications. However, the I$_{ON}$ at V$_{DD}$ = 0.5 V may be required to be enhanced further.

## V.  CONCLUSIONS

In this work, we have studied the implications of the application of a dual material gate in a TFET to simultaneously improve the overall performance of the device. We have demonstrated that the technique of using DMG can be applied in a DGTFET to obtain a good I$_{ON}$, I$_{OFF}$ and V$_T$ tradeoffs along with an improvement in the SS$_{AVG}$, improve the nature of the output characteristics and increase the immunity against the DIBL effects. We have also shown that DMG provides a technique to exploit the advantages of both a high and a low work-function gate in a TFET and thereby obviate the problems of both these types of gates. The improvement in device characteristics obtained using a DMG in an SDGTFET further demonstrate that the technique of DMG is capable of improving the overall device characteristics in the TFETs that are: a) using smaller band-gap channel material like sSOI b) based on high-k gate dielectric (HfO$_2$) c) having smaller t$_{ox}$=2 nm d) having smaller channel length L=25 nm e) driven by lower V$_{DD}$=0.5 V. It is worth mentioning that using a DMG in a TFET will bring further complexity to the device fabrication. Also, fabricating a DMG at a very small device dimension (channel length less than 100 nm) and extending it to below 10 nm may be challenging. However, since there is an appreciable improvement in the overall device characteristics, the complexity in the device fabrication might be acceptable and further fabrication-driven experiments on DMG at smaller dimensions are worth exploring.

Acknowledgment: This work was supported in part by the NXP (Philips) Chair Professorship awarded to M. Jagadesh Kumar.

REFERENCES

[1] Q. Zhang, W. Zhao, and A. Seabaugh, "Low-Subthreshold-Swing Tunnel Transistors", *IEEE Electron Device Letters*, vol. 27, pp. 297-300, April 2006.

[2] S. O. Koswatta, M. S. Lundstrom, and D.E. Nikonov, , "Performance Comparison Between p-i-n Tunneling Transistors and Conventional MOSFETs", *IEEE Trans. on Electron Devices*, vol. 56, pp. 456-465, March 2009.

[3] T. Nirschl, P.F. Wang, C. Weber, J. Sedlmeir, R. Heinrich, R. Kakoschke, K. Schrufer, J. Holz, C. Pacha, T. Schulz, M. Ostermayr, A. Olbrich, G. Georgakos, E. Ruderer, W. Hansch, and D. Schmitt-Landsiedel, "The Tunneling Field Effect Transistor (TFET) as an Add-on for Ultra-Low-Voltage Analog and Digital Processes", in *IEDM Tech. Digest*, pp. 195-198, Dec. 2004.

[4] K. K. Bhuwalka, J. Schulze, and I. Eisele, "Scaling the Vertical Tunnel FET with Tunnel Bandgap Modulation and Gate Work function Engineering", *IEEE Trans. Electron Devices*, vol. 52, pp. 909-917, May 2005.

[5] K. Boucart and A.M. Ionescu, "Double-Gate Tunnel FET with High-κ Gate Dielectric", *IEEE Trans. Electron Devices*, vol. 54, pp. 1725-1733, July 2007.

[6] E.H. Toh, G.H. Wang, L. Chan, G. Samudra, and Y.C. Yeo, "Device physics and guiding principles for the design of double-gate tunneling field effect transistor with silicon-germanium source heterojunction", *Applied Physics Letters*, vol. 91, Article Number: 243505, Dec. 2007.

[7] V.V. Nagavarapu, R.R. Jhaveri and J.C.S. Woo, "The Tunnel Source (PNPN) n-MOSFET: A Novel High Performance Transistor", *IEEE Trans. on Electron Devices*, vol. 55, pp.1013-1019, April 2008.

[8] A.S. Verhulst, W.G. Vandenberghe, K. Maex, S.D. Gendt, M.M. Heyns, and G. Groesenrken, "Complementary Silicon-Based Heterostructure Tunnel-FETs with High Tunnel Rates", *IEEE Electron Device Letters*, vol. 29, pp. 1398-1401, Dec. 2008.

[9] M. Schlosser, K.K. Bhuwalka, M. Sauter, T. Zilbauer, T. Sulima, and I. Eisele, "Fringing-Induced Drain Current Improvement in the Tunnel Field-Effect Transistor with High-k Gate Dielectrics", *IEEE Trans. on Electron Devices*, vol. 56, pp. 100-108, Jan. 2009.

[10]  S. Saurabh and M.J. Kumar, "Impact of Strain on Drain Current and Threshold Voltage of Nanoscale Double Gate Tunnel Field Effect Transistor: Theoretical Investigation and Analysis", *Japanese Journal of Applied Physics,* vol. 48, Article Number: 064503, Part 1, June 2009.

[11]  P. F. Guo, L. T. Yang, Y. Yang, L. Fan, G. Q. Han, G. S. Samudra, and Y. C. Yeo, "Tunneling Field-Effect Transistor: Effect of Strain and Temperature on Tunneling Current", *IEEE Electron Device Letters,* vol. 30, pp. 981-983, September 2009.

[12]  A. S. Verhulst, W. G. Vandenberghe, K. Maex, and G. Groeseneken, "Tunnel field-effect transistor without gate-drain overlap", *Applied Physics Letters*, vol. 91,  Article Number: 053102,  2007.

[13]  F. Mayer, C. L. Royer, J. F. Damlencourt, K. Romanjek, F. Andrieu, C. Tabone, B. Previtali, and S. Deleonibus, "Impact of SOI, Si1-xGexOI and GeOI substrates on CMOS compatible Tunnel FET performance", in *IEDM*, pp. 1-5, 2008

[14]  T. Krishnamohan, D. Kim, S. Raghunathan, and K. Saraswat, "Double-Gate Strained-Ge Heterostructure Tunneling FET (TFET) With Record High Drive Currents and <60mV/dec Subthreshold Slope", *IEDM*, pp. 1-3, Dec. 2008.

[15]  S. Saurabh and M.J. Kumar, "Estimation and Compensation of Process Induced Variations in Nanoscale Tunnel Field Effect Transistors (TFETs) for Improved Reliability", *IEEE Trans. Devices and Materials Reliability,* Vol.10, pp.390 - 395, September 2010.

[16]  C. Shen, S.L. Ong, C.H. Heng, G. Samudra, and Y.C. Yeo, "A Variational Approach to the Two-Dimensional Nonlinear Poisson's Equation for the Modeling of Tunneling Transistors", *IEEE Electron Device Letters,* vol. 29, pp. 1252-1255, Nov. 2008.

[17]  K. Boucart  and A. M. Ionescu, "Length scaling of the Double Gate Tunnel FET with a high-K gate dielectric", *Solid-State Electronics*, vol. 51, pp. 1500–1507, Nov. 2007

[18]  K. Boucart and A.M. Ionescua, "A new definition of threshold voltage in Tunnel FETs"*, Solid-State Electronics,* vol. 52, pp. 1318-1323, Sept. 2008.

[19]  Y. Khatami and K. Banerjee, "Steep Subthreshold Slope n- and p-Type Tunnel-FET Devices for Low-Power and Energy-Efficient Digital Circuits", *IEEE Trans. on Electron Devices*, vol. 56, pp. 2752-2761, Nov. 2009.

[20]  Semiconductor Industry Association (SIA), International Technology Roadmap for Semiconductors, 2009 Edition.

[21]  W. Long, H. Ou, J.M. Kuo, and K.K. Chin, "Dual-Material Gate (DMG) Field Effect Transistor", *IEEE Trans. Electron Devices,* vol. 46, pp. 865-870, May 1999.

[22]  K.Y. Na and Y.S. Kim, "Silicon Complementary Metal–Oxide–Semiconductor Field-Effect Transistors with Dual Work Function Gate", *Japanese Journal of Applied Physics,* vol. 45, pp. 9033-9036, Dec. 2006.

[23]  A. Chaudhary and M.J.Kumar, "Investigation of the novel attributes of a fully depleted dual-material gate SOI MOSFET", *IEEE Trans. Electron Devices,* vol. 51, pp. 1463-1467, September 2004.

[24]  *Atlas User's Manual: Device Simulation Software*, Silvaco Int., Santa Clara, CA, 2008.

[25]  K. Rim, K. Chan, L. Shi, D. Boyd, J. Ott, N. Klymko, F. Cardone, L. Tai, S. Koester, M. Cobb, D. Canaperi, B. To, E. Duch, I. Babich, R. Carruthers, P. Saunders, G. Walker, Y. Zhang, M. Steen, and M. Ieong, "Fabrication and mobility characteristics of ultra-thin strained Si directly on Insulator (SSDOI) MOSFETs", in *IEDM Tech. Digest*, pp. 3.1.1-3.1.4, 2003.

[26]  S.Takagi , T. Mizuno, T. Tezuka, N. Sugiyama, T. Numata, K. Usuda, Y. Moriyama, S. Nakaharai, J. Koga, A. Tanabe,q and T. Maeda, "Fabrication and device characteristics of strained-Si-on-insulator (strained SOI) CMOS", *Applied Surface Science*, vol. 224, pp. 241-247, March 2004.

[27]  T. Numata, T. Mizuno, T. Tezuka, J. Koga, and S. Takagi, "Control of threshold-voltage and short-channel effects in ultrathin strained-SOI CMOS devices", *IEEE Trans. on Electron Devices*, vol. 52, pp. 1780-1786, Aug. 2005.

[28]  P. Ranade, Y.C. Yeo, Q. Lu, H. Takeuchi, T.J. King, and C. Hu, "Molybdenum As A Gate Electrode For Deep Sub-Micron CMOS Technology", *MRS Symposium,* vol. 611, C3.2.1-6, (2000).

[29]  I. Polishchuk, P. Ranade, T.J. King, and C. Hu, "Dual Work Function Metal Gate CMOS Transistors by Ni–Ti Interdiffusion", *IEEE Electron Device Letters,* vol. 23, pp. 200-202, April 2002.

[30]  M. Hasan, H. Park, H. Yang, H. Hwang, H.S. Jung, and J.H. Lee, "Ultralow work function of scandium metal gate with tantalum nitride interface layer for n-channel metal oxide semiconductor application n-channel metal oxide semiconductor application", *Applied Physics Letters,* vol. 90, pp. 103510 - 103510-3, March 2007.

[31]  W. Y. Choi and W. Lee, "Hetero-Gate-Dielectric Tunneling Field Effect Transistors", *IEEE Trans. on Electron Devices*, vol. 57, pp. 2317-2319, Sept. 2010.

# List of Figures

Fig. 11: Transfer characteristics of the DMG-SDGTFET ($\Phi_{tunnel}$ = 4.5 eV, $\Phi_{aux}$ = 4.7 eV, x = 0.5, $L_{tunn}$ = 10 nm , $L_{aux}$ = 15 nm and $V_{DS}$ = 0.5 V) and SMG-SDGTFET (.......... $\Phi_m$ = 4.5 eV ▬ ▬ $\Phi_m$ = 4.7 eV, x = 0.5, L = 25 nm and $V_{DS}$ = 0.5 V).

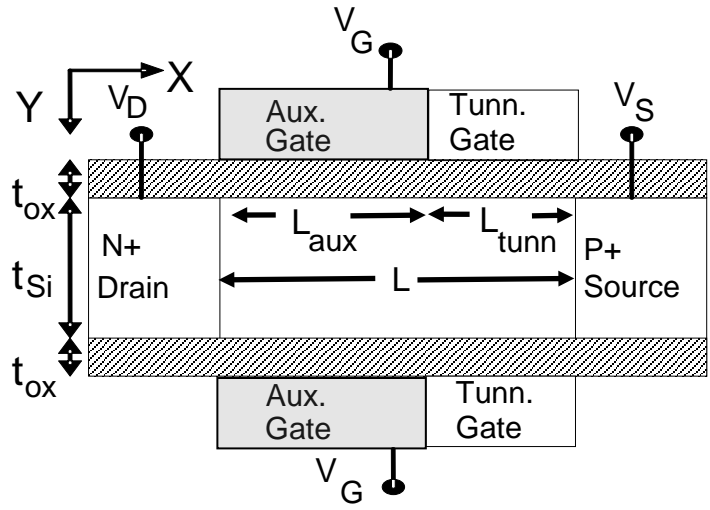

**Fig. 1**

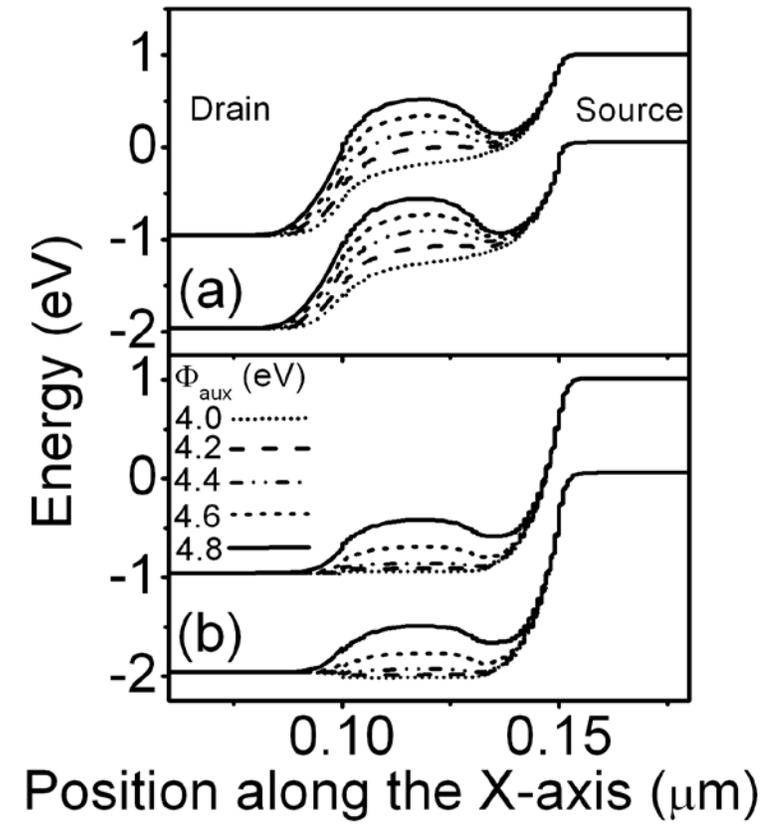

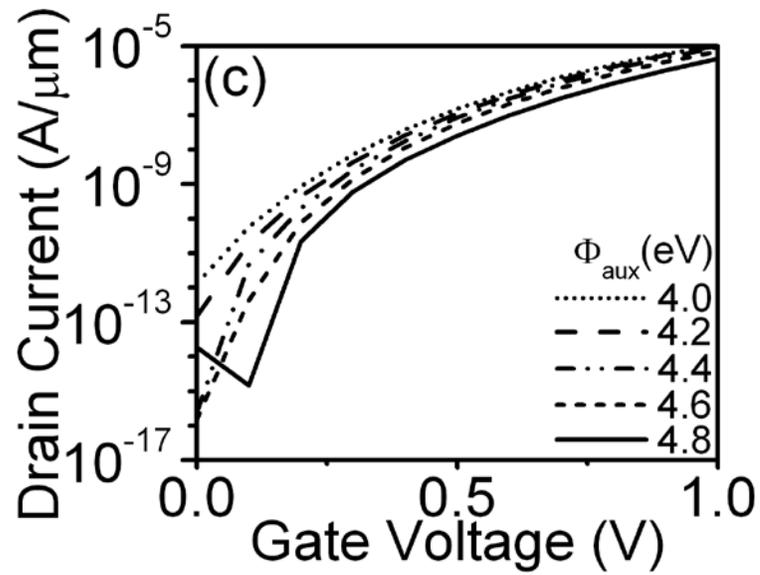

**Fig. 2**

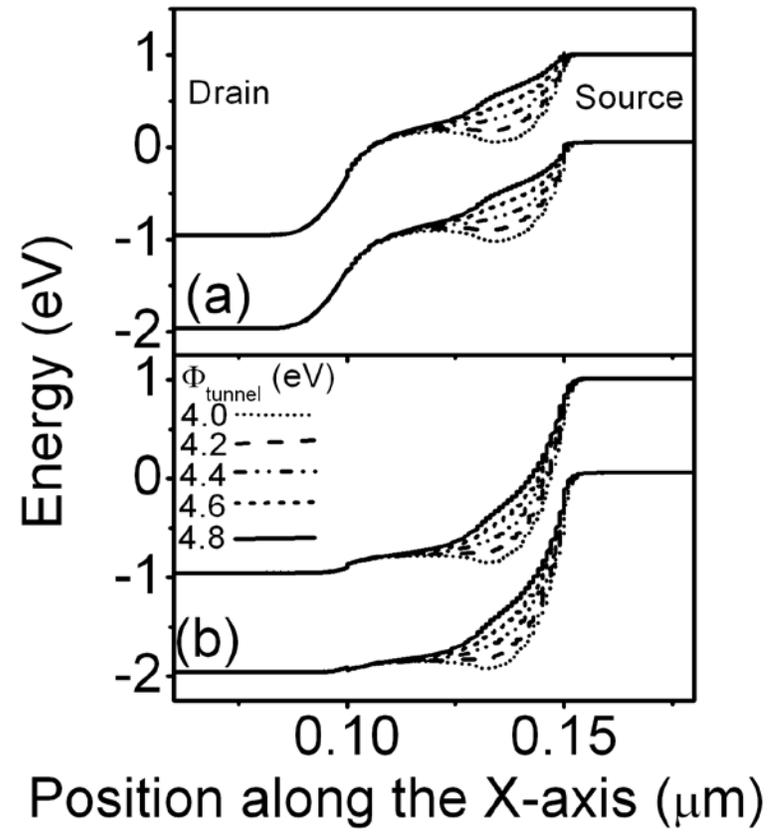

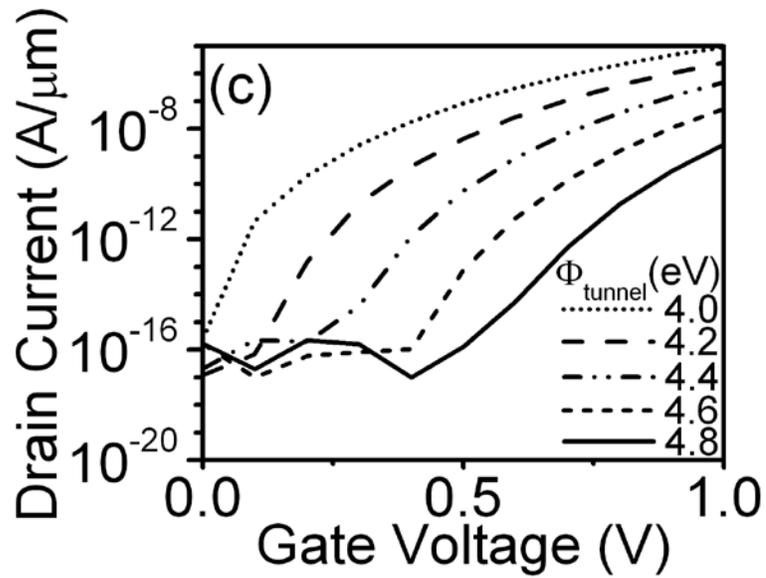

**Fig. 3**

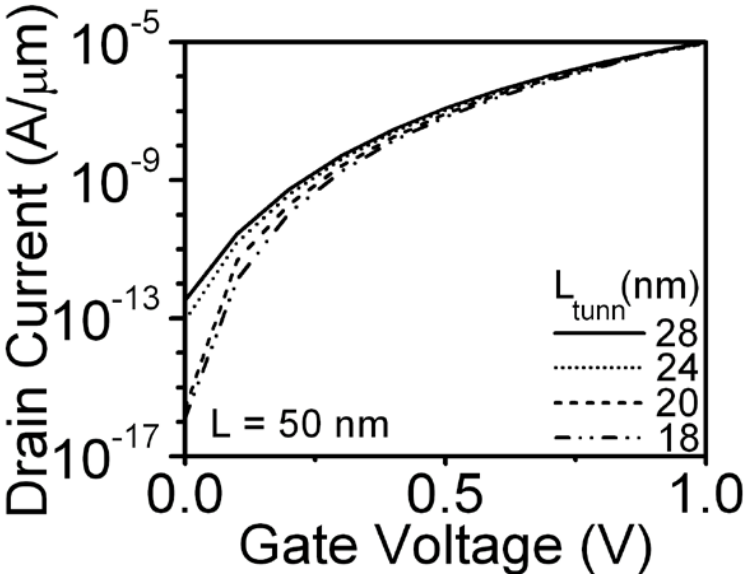

**Fig. 4**

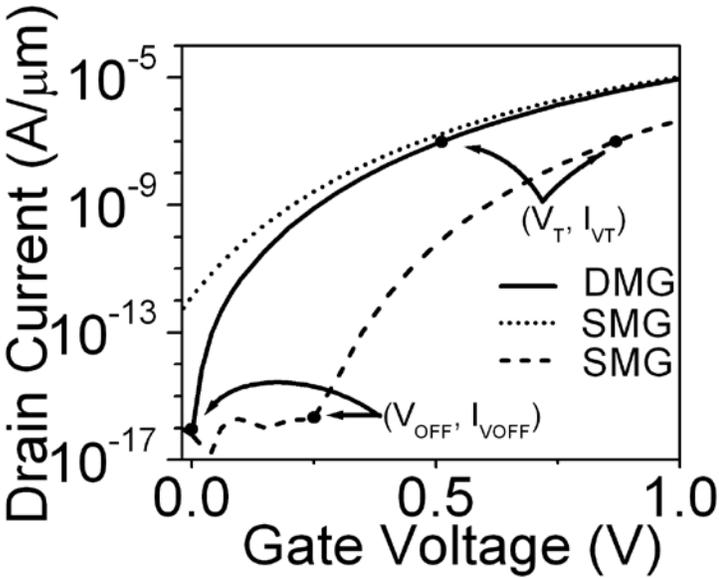

**Fig. 5**

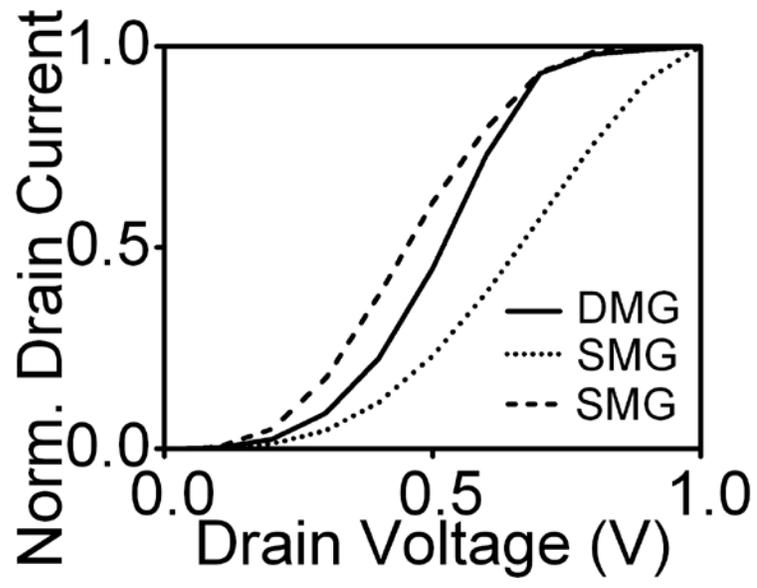

**Fig. 6**

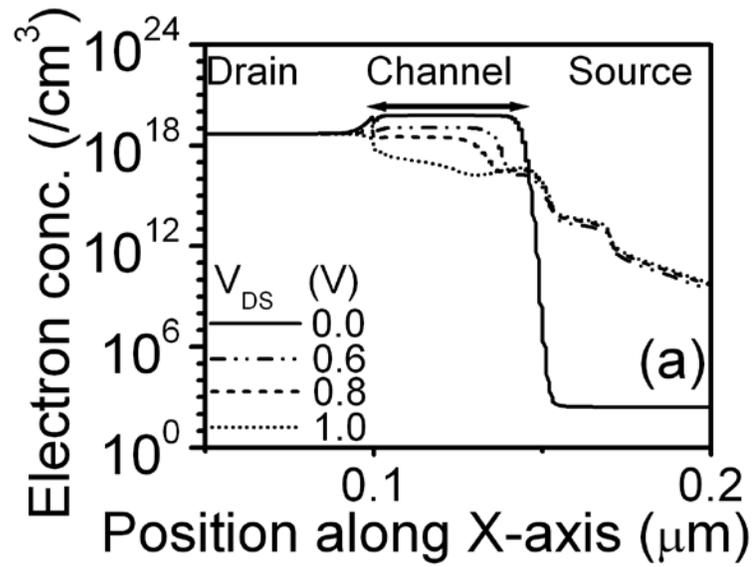

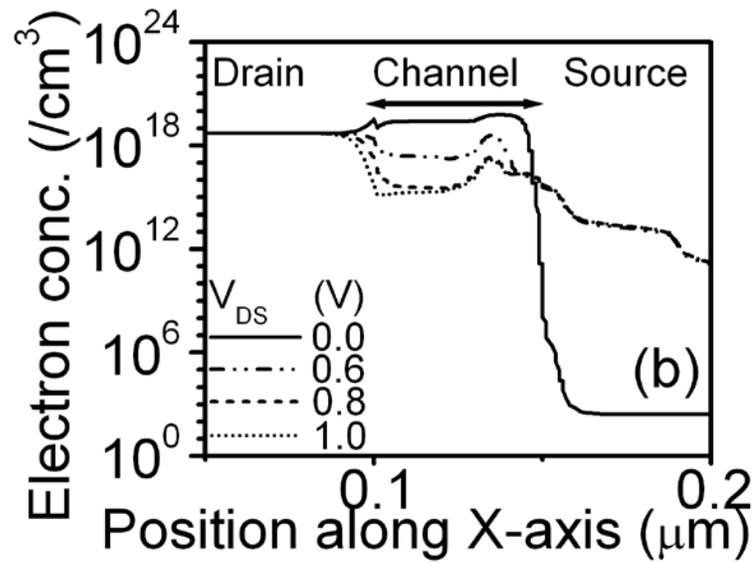

**Fig. 7**

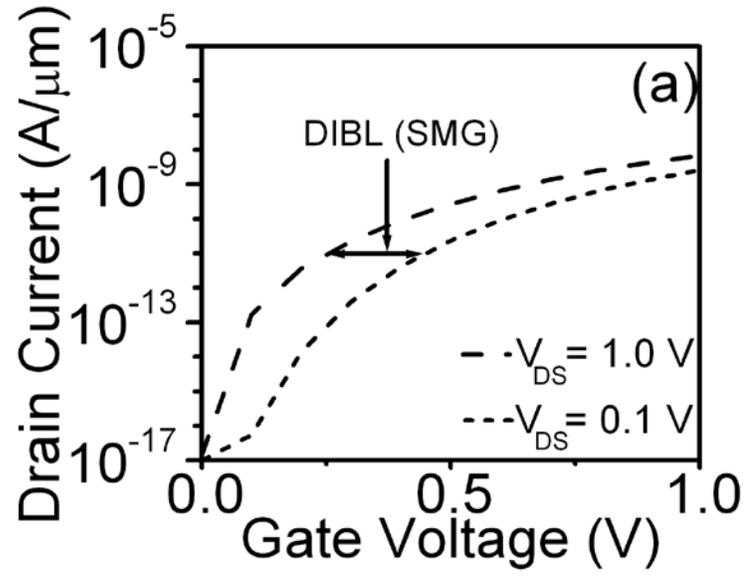

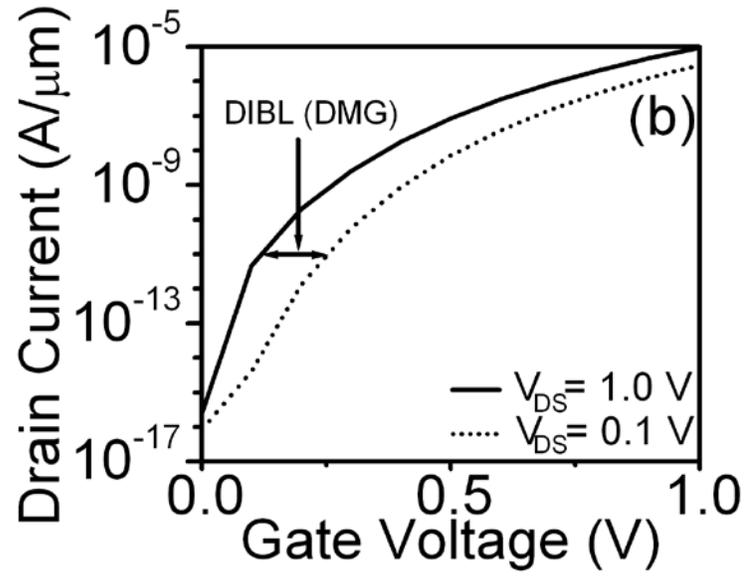

**Fig. 8**

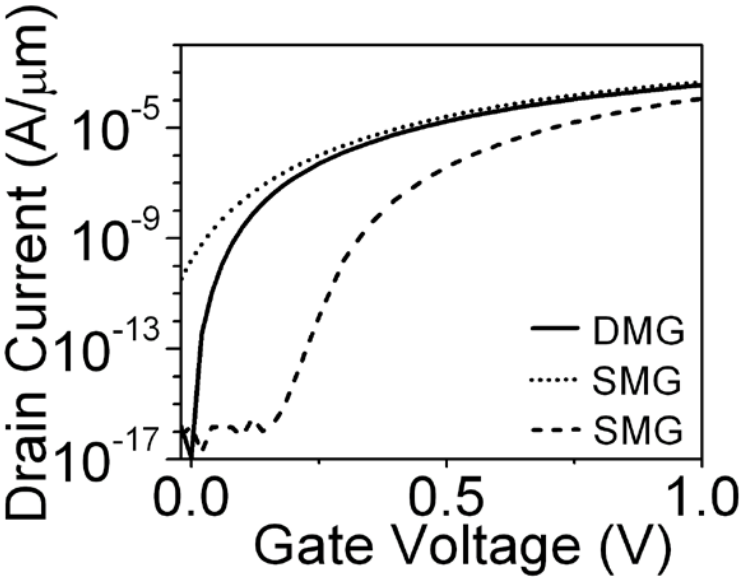

**Fig. 9**

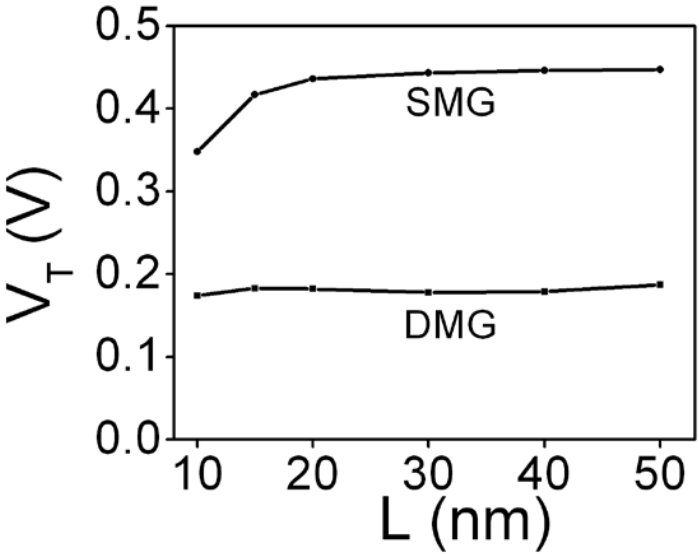

**Fig. 10**

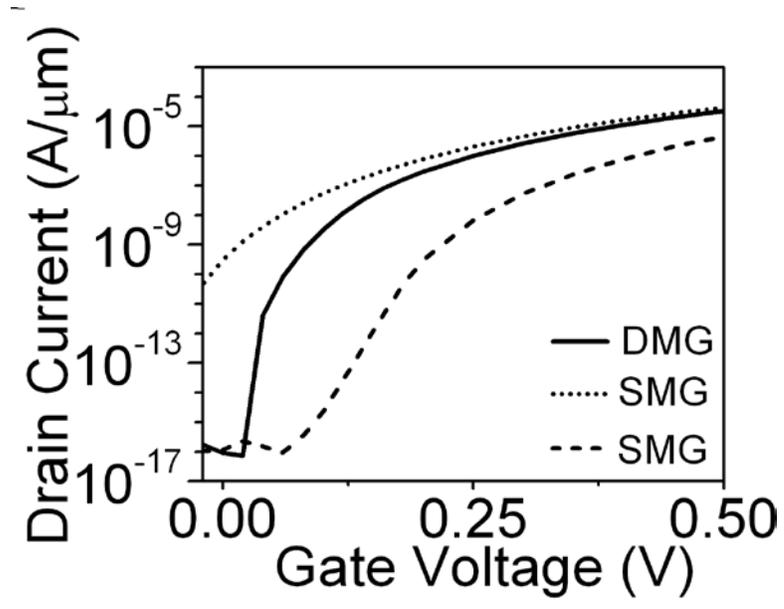

**Fig. 11**

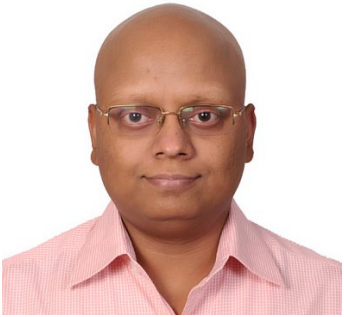

**Sneh Saurabh** received the B. Tech. degree in Electrical Engineering from Indian Institute of Technology, Kharagpur in 2000. Since then he has been working on various aspects of design and verification of integrated circuits. His research interests include quantum devices, low-power methodologies and design optimizations. He is currently pursuing his PhD at Indian Institute of Technology, New Delhi in the field of tunnel devices for CMOS applications.

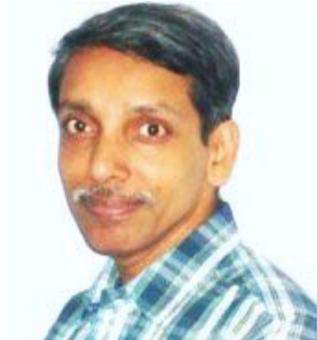

M. Jagadesh Kumar was born in Mamidala, Andhra Pradesh, India. He received the M.S. and Ph.D. degrees in electrical engineering from the Indian Institute of Technology (IIT), Madras, India.

From 1991 to 1994, he performed a postdoctoral research on the modeling and processing of highspeed bipolar transistors with the Department of Electrical and Computer Engineering, University of Waterloo, Waterloo, ON, Canada. While with the University of Waterloo, he also did research on amorphous-silicon thin-film transistors. From July 1994 to December 1995, he was initially with the Department of Electronics and Electrical Communication Engineering, IIT, Kharagpur, India, and then, he was with the Department of Electrical Engineering, IIT, New Delhi, India, where he became an Associate Professor in July 1997 and has been a Full Professor in January 2005. He is currently the Chair Professor of the NXP (Philips) (currently, NXP Semiconductors India Pvt. Ltd.) established at IIT Delhi by Philips Semiconductors, The Netherlands. He is the Coordinator of the Very Large Scale Integration (VLSI) Design, Tools, and Technology interdisciplinary program at IIT Delhi.

His research interests include nanoelectronic devices, device modeling and simulation for nanoscale applications, integrated-circuit technology, and power semiconductor devices. He has published extensively in these areas of research with three book chapters and more than 145 publications in refereed journals and conferences. His teaching has often been rated as outstanding by the Faculty Appraisal Committee, IIT Delhi.

Dr. Kumar is a fellow of the Indian National Academy of Engineering, The National Academy of Sciences and the Institution of Electronics and Telecommunication Engineers (IETE), India. He is recognized as a Distinguished Lecturer of the IEEE Electron Devices Society (EDS). He is a member of the EDS Publications Committee and the EDS Educational Activities Committee. He is an Editor of the IEEE TRANSACTIONS ON ELECTRON DEVICES. He was the lead Guest Editor for the following: 1) the joint special issue of the IEEE TRANSACTIONS ON ELECTRON DEVICES and the IEEE TRANSACTIONS ON NANOTECHNOLOGY (November 2008 issue) on Nanowire Transistors: Modeling, Device Design, and Technology and 2) the special issue of the IEEE TRANSACTIONS ON ELECTRON DEVICES on Light Emitting Diodes (January 2010 issue). He is the Editor-in-Chief of the IETE Technical Review and an Associate Editor of the Journal of Computational Electronics. He is also on the editorial board of Recent Patents on Nanotechnology, Recent Patents on Electrical Engineering, Journal of Low Power Electronics, and Journal of Nanoscience and Nanotechnology. He has reviewed extensively for different international journals.

He was a recipient of the 29th IETE Ram LalWadhwa GoldMedal for his distinguished contribution in the field of semiconductor device design and modeling. He was also the first recipient of the India Semiconductor Association–VLSI Society of India TechnoMentor Award given by the India Semiconductor Association to recognize a distinguished Indian academician for playing a significant role as a Mentor and Researcher. He is also a recipient of the 2008 IBM Faculty Award. He was the Chairman of the Fellowship Committee of The Sixteenth International Conference on VLSI Design (January 4–8, 2003, New Delhi, India), the Chairman of the Technical Committee for High Frequency Devices of the International Workshop on the Physics of Semiconductor Devices (December 13–17, 2005, New Delhi), the Student Track Chairman of the 22nd International Conference on VLSI Design (January 5–9, 2009, New Delhi), and the Program Committee Chairman of the Second International Workshop on Electron Devices and Semiconductor Technology (June 1–2, 2009, Mumbai, India).